\documentclass[aps, pra, twocolumn, amsmath, amssymb, 
superscriptaddress, 
nofootinbib]{revtex4-1}

\usepackage[utf8]{inputenc}
\usepackage{graphicx}
\usepackage{units}
\usepackage{color}
\usepackage[pdftex, colorlinks=true, linkcolor=myblue, citecolor=myblue,
urlcolor=myblue]{hyperref}
\usepackage{comment}

\definecolor{myblue}{rgb}{0,0,1}

\begin{document}

\title{Retardation effects on the dispersion and propagation of plasmons in metallic nanoparticle chains}

\author{Charles A. Downing}
\altaffiliation{Present address: Departamento de F\'isica Te\'orica de la Materia Condensada and Condensed Matter Physics Center (IFIMAC), Universidad 
Aut\'onoma de Madrid, E- 28049 Madrid, Spain}
\affiliation{Universit\'{e} de Strasbourg, CNRS, Institut de Physique et Chimie des Mat\'{e}riaux de Strasbourg,  UMR 7504, F-67000 Strasbourg, France}

\author{Eros Mariani}
\affiliation{School of Physics and Astronomy, University of Exeter, Stocker Rd.\ EX4 4QL Exeter, UK}

\author{Guillaume Weick}
\email{guillaume.weick@ipcms.unistra.fr} 
\affiliation{Universit\'{e} de Strasbourg, CNRS, Institut de Physique et Chimie des Mat\'{e}riaux de Strasbourg,  UMR 7504, F-67000 Strasbourg, France}


\begin{abstract}
We consider a chain of regularly-spaced spherical metallic nanoparticles, where each particle supports three degenerate localized surface plasmons. Due to the dipolar interaction between 
the nanoparticles, the localized plasmons couple to form extended collective modes. 
Using an open quantum system approach in which the collective plasmons are interacting with vacuum electromagnetic modes and which, importantly, readily incorporates retardation via the light-matter coupling, we analytically evaluate the resulting radiative frequency shifts of the plasmonic bandstructure. 
For subwavelength-sized nanoparticles, our analytical treatment provides an excellent quantitative agreement with the results stemming from
laborious numerical calculations based on fully-retarded solutions to Maxwell's equations. 
Indeed, the explicit expressions for the plasmonic spectrum which we provide showcase how including retardation gives rise to a logarithmic singularity in the bandstructure of transverse-polarized plasmons.
We further study the impact of retardation effects on the propagation of plasmonic excitations along the chain. While for the longitudinal modes, retardation has a negligible effect, we find that the retarded dipolar interaction can significantly modify the plasmon propagation in the case of transverse-polarized modes.  
Moreover, our results elucidate the analogy between radiative effects in nanoplasmonic systems and the cooperative Lamb shift in atomic physics. 
\end{abstract}

\maketitle

\section{Introduction}
\label{Sec:intro}

The ability to confine and control light at the nanoscale is a major achievement of plasmonic systems \cite{Ozbay2006, Gramotnev2010, Stockman2011}. It is expected that such an appealing
 property will allow the field to spawn numerous applications, in areas ranging from subwavelength optics and data storage to light generation, microscopy and biophotonics \cite{Barnes2003}.

The plasmonic quasiparticle, a collective oscillation of electrons in a metal, may occur in a variety of different forms, such as a bulk plasmon inside the volume of a metallic solid, a surface plasmon at a metal-dielectric interface, or a localized surface plasmon (LSP) in a metallic nanoparticle \cite{Maier2007}.

Nearly twenty years ago, it was suggested \cite{Quinten1998} that a linear chain of regularly-spaced metal nanoparticles could be used as a subwavelength light guide by exploiting plasmons. The idea was to harness the electrodynamic interparticle coupling between the LSPs to transmit light along the effective waveguide.  Such a system is thought to be a key component in future plasmonic circuitry, and consequently there has been a plethora of pioneering experimental \cite{Krenn1999, Maier2002, Maier2003a, Koendrick2007, Crozier2007, Apuzzo2013, Barrow2014} and theoretical 
\cite{Brongersma2000, Maier2003b, Park2004, Weber2004, Citrin2004, Simovski2005, Citrin2006, Koenderink2006, Markel2007, Fung2007, Petrov2015, Lee2012, Pino2014, Brandstetter2016} investigations seeking to achieve energy and information transport over macroscopic distances using metallic nanoparticle chains. Furthermore, bipartite chains have also been shown to be of fundamental interest due to their inherent topologically-nontrivial behavior \cite{Poddubny2014, Ling2015, Bipartite2017, pococ17_preprint}.

The electrodynamic interparticle interaction in the chain leads to coupling of the LSPs into collective plasmonic excitations that are extended over the whole one-dimensional array. This in turn gives rise to a collective plasmonic bandstructure, in direct analogy with the quasiparticles (like electrons or phonons) encountered in one-dimensional lattices in solid-state physics. A number of early theoretical investigations sought to map out the plasmonic bandstructure, predominantly from numerical solutions to Maxwell's equations in the quasistatic limit, i.e., without including the effects of retardation in the far field~\cite{Quinten1998, Brongersma2000, Maier2003b, Park2004}. 

It was first noticed by Weber and Ford \cite{Weber2004}, and independently by Citrin \cite{Citrin2004}, that retardation has a significant impact on the plasmonic bandstructure 
and results in radiative shifts of the quasistatic dispersion relation. In particular, a cusp was found to appear at the intersection of the quasistatic plasmonic spectrum with the light cone for the case of transverse-polarized plasmons (i.e., those with dipole moments pointing perpendicular to the chain). 
Subsequently, further investigations \cite{Simovski2005, Citrin2006, Koenderink2006, Markel2007, Fung2007, Petrov2015}, principally using numerical and semi-analytical solutions to Maxwell's equations, confirmed the results of 
Refs.~\cite{Weber2004, Citrin2004} and the importance of retardation.  However, for these aforementioned calculations to be fully consistent, they require the introduction of a correction to the polarizability due to radiation damping, a term which is not universally agreed upon in the literature and which can lead to acausal behavior  \cite{Weber2004}.
Furthermore, it is well-known that the frequency shift for a single radiating oscillator, as obtained classically with the introduction of an Abraham-Lorentz term in the equation of motion, is an order of magnitude smaller than the result arising from a quantum mechanical calculation \cite{Jackson}. 
Notably, in the pioneering experiments of Lamb and Retherford on the fine structure of the hydrogen atom \cite{Lamb1947}, it is the quantum theory \cite{Bethe1947} which describes the experimental data with spectacular quantitative agreement.

More recently, several groups also developed quantum treatments of plasmonic chains \cite{Lee2012, Pino2014, Brandstetter2016}, neglecting retardation effects. A quantum approach is particularly needed when the size of the nanoparticles constituting the chain is such that quantum-size effects are important~\cite{Tame2013}.
Moreover, as we show below, theoretical tools borrowed from quantum optics provide a useful and straightforward framework for investigating retardation effects in nanoplasmonic systems. 

In this work, we use an open quantum system approach to systematically study retardation effects on the collective plasmon bandstructure in chains of metallic nanoparticles in a fully self-contained manner. 
Such an approach allows one to directly access the finite plasmonic lifetimes~\cite{Brandstetter2016}, which arise due to the irreversible dissipation of energy from the plasmonic system to the three-dimensional photonic bath to which it is coupled. 
As follows from the fluctuation-dissipation theorem \cite{Kubo1966}, the photonic environment further gives rise to a shift in the plasmonic energy levels \cite{Downing2017}, in direct analogy with the celebrated Lamb shift in atomic physics \cite{Bethe1947, Milonni1994}. 
Advantageously, our approach allows us to uncover simple analytical expressions which provide unique insight into the phenomena under investigation. 
In particular, we reveal for long chains that the aforementioned cusp in the transverse-polarized plasmonic bandstructure corresponds to a logarithmic singularity.
We further perform numerical calculations based on the fully-retarded solutions to Maxwell's equations for a finite chain of point dipoles~\cite{Weber2004} and find 
excellent quantitative agreement with our analytical theory in the limit of small nanoparticles. 

Our open quantum system approach also allows us to study the transport of plasmonic excitations along the chain, and to elucidate the effect of retardation on it. We find that such effects, as well as the long-ranged nature of the dipole-dipole interaction, are unimportant (at the qualitative level) for the transport of longitudinal plasmonic excitations. In contrast, retardation effects can have an important impact on electromagnetic energy transport in the case of transverse excitations. 

The radiative frequency shifts studied here are connected to the so-called cooperative Lamb shift, familiar from many-atom systems, where an enhancement of the Lamb shift due to collective interactions between particles is exhibited \cite{Friedberg1973, Scully2009}.
Such a cooperative Lamb shift has been measured in a variety of pioneering atomic physics experiments \cite{Garrett1990, Rohlsberger2010, Keaveney2012, Meir2014}, including most recently a synthetic vacuum using ultracold atomic gas mixtures~\cite{Rentrop2016}. One consequence of our work is the proposal that the experimental detection of radiative shifts in a chain of metallic nanoparticles would constitute a realization of the cooperative Lamb shift in nanoplasmonics.

The paper is organized as follows: In Sec.~\ref{Sec:model}, we present our model of a chain of plasmonic nanoparticles coupled to vacuum photonic modes. In Sec.\ \ref{sec:qsa} we derive the quasistatic plasmonic bandstructure. We unveil analytical expressions for the radiative shifts of the collective plasmonic bandstructure in Sec.~\ref{Sec:shifts} and compare them to classical electrodynamic numerical calculations in Sec.~\ref{sec:numerics}. 
In Sec.\ \ref{sec:propagation}, we study the influence of retardation effects on the transport of plasmonic excitations along the chain. 
Finally, we draw conclusions in Sec.~\ref{sec:conc}.
The appendix presents our analytical result for the radiative decay rates of the collective plasmons, including the long-ranged dipole-dipole interaction, and compares such a result to electromagnetic numerical calculations.

\section{Model}
\label{Sec:model}
We start by presenting our model, which builds upon the quantum theory developed in Ref.~\cite{Brandstetter2016}, with the significant extensions of including the effects of both long-range quasistatic interactions and retardation.
We note that a classical model could also be used (cf.\ Sec.\ \ref{sec:numerics}), with the significant cost of losing complete integrability.

\begin{figure}[tb]
 \includegraphics[width=\linewidth]{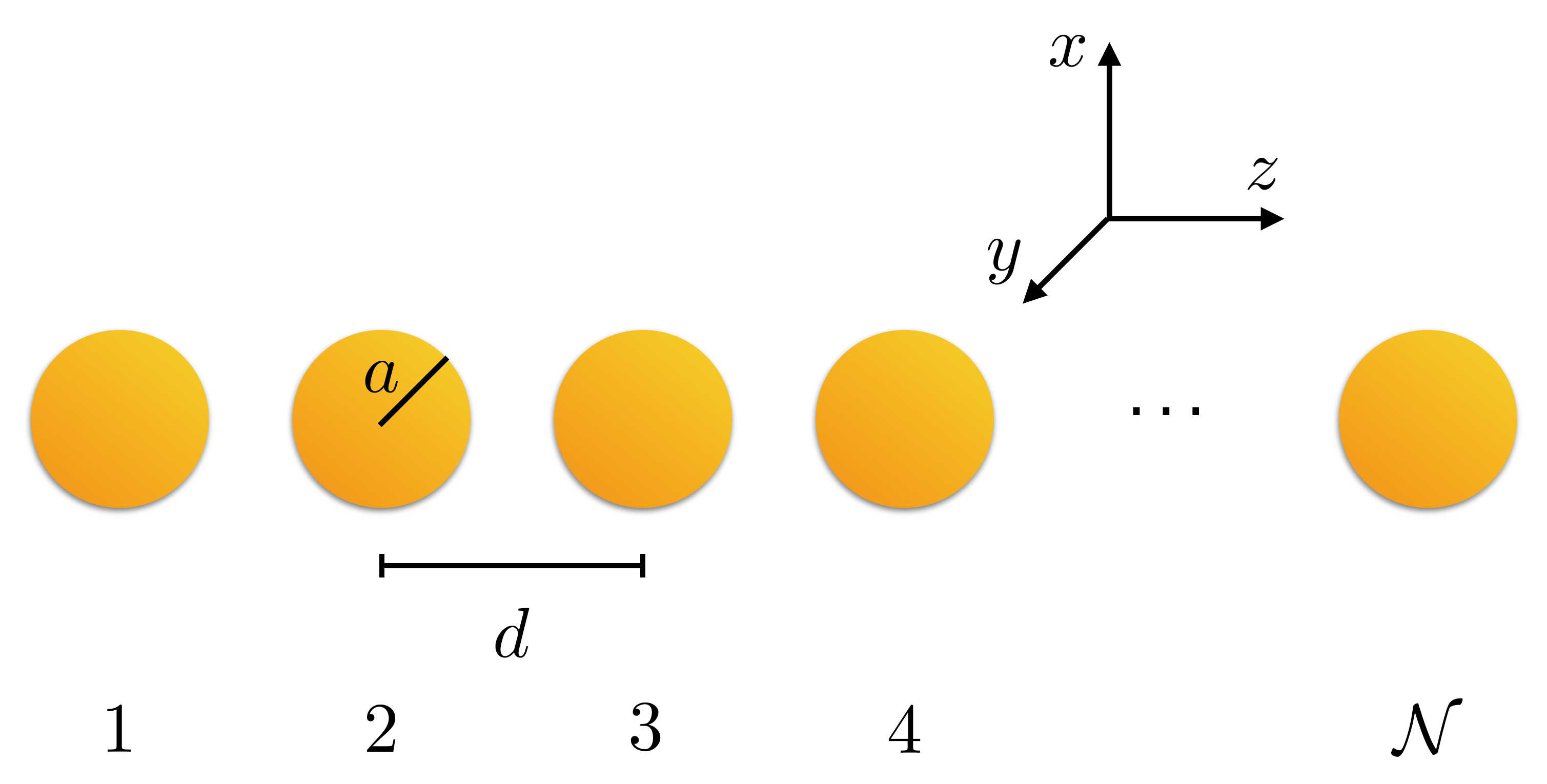}
 \caption{Sketch of a chain of $\mathcal{N}$ spherical metallic nanoparticles of radius $a$ separated by a center-to-center distance $d$.}
 \label{fig:sketch}
\end{figure}

Specifically, we consider a one-dimensional array of spherical metallic nanoparticles of radius $a$ separated by a center-to-center distance $d$ (see Fig.\ \ref{fig:sketch}). Each nanoparticle contains $N_\mathrm{e}$ valence electrons of charge $-e<0$ and mass $m_\mathrm{e}$ and supports three degenerate, orthogonal dipolar LSPs polarized along the directions $x$, $y$ or $z$. Each LSP corresponds to a harmonic oscillation of the electronic center of mass at the (bare) resonance frequency $\omega_0$. The latter quantity coincides with the Mie frequency $\omega_{\mathrm{p}} / \sqrt{3} = \sqrt{N_\mathrm{e} e^2/m_\mathrm{e} a^3}$ for the case of alkaline nanoparticles in vacuum, where $\omega_{\text{p}}$ is the plasma frequency. Coulomb interactions between the LSPs in the chain, which are essentially dipolar for an interparticle separation $d\gtrsim 3a$~\cite{Park2004}, lead to the coupling of the localized plasmonic modes into collective plasmons extended over the whole chain. In the following, we show that the resulting 
collective plasmonic bandstructure is highly modified by retardation effects, accounted for in our model by the coupling of the plasmonic modes to the three-dimensional photonic environment. 

Throughout this work we neglect the effects of Landau damping \cite{Kreibig, Bertsch, Kawabata1966, Weick2005}, and the associated shift it induces in the plasmonic resonance frequency \cite{Weick2006}, since we are primarily focused on radiative effects, which are dominant as long as the nanoparticles are not too small (i.e., their radius should be more than ca.\ \unit[5]{nm}).

In the Coulomb gauge \cite{cohen, craig}, the fully-retarded Hamiltonian of the plasmonic chain coupled to vacuum electromagnetic modes in a 
volume $\mathcal{V}$ reads 
\begin{equation}
\label{eq:Ham}
H=H_{\mathrm{pl}} + H_\mathrm{ph} + H_{\mathrm{pl}\textrm{-}\mathrm{ph}}.
\end{equation}
The purely plasmonic Hamiltonian describing the LSPs coupled through the long-ranged quasistatic dipole-dipole interaction
is \cite{Weick2013, Brandstetter2015}
\begin{align}
\label{eq:Ham_chain}
  H_{\mathrm{pl}} =&\; \sum_{\sigma=x,y,z}\bigg[\hbar \omega_0 \sum_{n=1}^\mathcal{N} {b_{n}^{\sigma}}^{\dagger} b_{n}^{\sigma} 
  \nonumber\\
&+ \frac{\hbar \Omega}{2} \sum_{\substack{n,m=1\\(n\neq m)}}^\mathcal{N} \frac{\eta_{\sigma}}{|n-m|^3} \left( b_{n}^{\sigma} + {b_{n}^{\sigma}}^{\dagger} \right) \left( b_{m}^{\sigma} + {b_{m}^{\sigma}}^{\dagger} \right)\bigg].
 \end{align}
Here, the indices $\{n,m\} \in [1, \mathcal{N}]$ denote the particle number in the chain of $\mathcal{N}$ nanoparticles
and $\sigma$ accounts for the two transverse ($x, y$) and the single longitudinal ($z$) polarizations of the plasmonic modes 
(see Fig.\ \ref{fig:sketch}).
The bosonic operator $b_{n}^{\sigma}$ (${b_{n}^{\sigma}}^{\dagger}$) annihilates (creates) an LSP with polarization $\sigma$ on nanoparticle $n$. The coupling constant is $\Omega = (\omega_0/2)({a}/{d})^3$, and the polarization-dependent factor $\eta_{x,y} = 1$ ($\eta_{z} = -2$) for the transverse (longitudinal) modes arises from the anisotropy of the dipolar interaction.

The photonic environment in Eq.\ \eqref{eq:Ham} is described by the Hamiltonian
\begin{equation}
\label{eq:H_ph}
 H_{\mathrm{ph}} = \sum_{\mathbf{k}, \hat{\lambda}_{\mathbf{k}}} \hbar \omega_{\mathbf{k}} {a_{\mathbf{k}}^{\hat{\lambda}_{\mathbf{k}}}}^{\dagger} a_{\mathbf{k}}^{\hat{\lambda}_{\mathbf{k}}}, 
\end{equation}
where $a_{\mathbf{k}}^{\hat{\lambda}_{\mathbf{k}}}$ (${a_{\mathbf{k}}^{\hat{\lambda}_{\mathbf{k}}}}^{\dagger}$) annihilates (creates) a photon with wavevector $\mathbf{k}$, transverse polarization 
 $\hat{\lambda}_{\mathbf{k}}$ (i.e., $\mathbf{k}\cdot\hat{\lambda}_{\mathbf{k}}=0$), and dispersion $\omega_{\mathbf{k}} = c |\mathbf{k}|$, where $c$ is the speed of light in vacuum. 

The plasmon-photon coupling Hamiltonian in Eq.~\eqref{eq:Ham} reads in the long-wavelength approximation ($k_0a\ll1$, with $k_0=\omega_0/c$) as
\cite{cohen, craig}
\begin{equation}
\label{eq:HamCoupling}
 H_{\mathrm{pl}\textrm{-}\mathrm{ph}} = \frac{e}{m_{\mathrm{e}}} \sum_{n=1}^{\mathcal{N}} \mathbf{\Pi}_n \cdot \mathbf{A} (\mathbf{d}_n) + \frac{N_{\mathrm{e}} e^2}{2 m_{\mathrm{e}}} \sum_{n=1}^{\mathcal{N}} \mathbf{A}^2 (\mathbf{d}_n), 
\end{equation}
where $\mathbf{d}_n = d (n-1) \hat{z}$ corresponds to the location of the center of nanoparticle $n$ (here and in what follows, hats designate unit vectors). The momentum associated with the LSPs in nanoparticle $n$ is 
\begin{equation}
\mathbf{\Pi}_n=\mathrm{i}\sqrt{\frac{N_\mathrm{e}m_\mathrm{e}\hbar\omega_0}{2}}\sum_{\sigma=x,y,z} \hat\sigma~({b_n^\sigma}^\dagger-b_n^\sigma),
\end{equation}
while the vector potential is given by
\begin{equation}
\label{eq:A}
\mathbf{A}(\mathbf{d}_n)=\sum_{\mathbf{k}, \hat\lambda_{\mathbf{k}}}
\hat\lambda_{\mathbf{k}}\sqrt{\frac{2\pi\hbar}{\mathcal{V}\omega_\mathbf{k}}}
\left(
a_\mathbf{k}^{\hat\lambda_{\mathbf{k}}}\mathrm{e}^{\mathrm{i}\mathbf{k}\cdot \mathbf{d}_n}
+{a_\mathbf{k}^{\hat\lambda_{\mathbf{k}}}}^\dagger\mathrm{e}^{-\mathrm{i}\mathbf{k}\cdot \mathbf{d}_n}
\right).
\end{equation}
Importantly, the Hamiltonian \eqref{eq:HamCoupling} fully takes into account retardation effects. In particular, the first term on the right-hand side of Eq.\ \eqref{eq:HamCoupling}, together with the quasistatic interaction in Eq.\ \eqref{eq:Ham_chain}, correspond to the retarded dipole-dipole interaction, as can be 
readily checked from second-order perturbation theory \cite{craig}. 
In Sec.\ \ref{Sec:shifts}, by means of second-order perturbation theory we show that the light-matter coupling in Eq.~\eqref{eq:HamCoupling} leads to the radiative frequency shifts which we evaluate analytically.

\section{Quasistatic plasmonic bandstructure}
\label{sec:qsa}

Before analyzing the effect of the photonic environment on the collective plasmon dispersion in Sec.\ \ref{Sec:shifts}, here we consider first the purely plasmonic Hamiltonian \eqref{eq:Ham_chain} and derive its associated quasistatic spectrum.

In the long-chain limit where $\mathcal{N} \gg 1$ \cite{footnote:infinite}, it is convenient to use periodic boundary conditions and to move into wavevector space via the Fourier transform 
$ b_{n}^{\sigma} = \mathcal{N}^{-1/2} \sum_{q} \mathrm{e}^{\mathrm{i} n q d}\, b_{q}^{\sigma}$, 
where the plasmonic wavevector $q =  2 \pi p /\mathcal{N} d$, with the integer $p\in[-\mathcal{N}/2, \mathcal{N}/2]$.
Then one obtains for the plasmonic Hamiltonian \eqref{eq:Ham_chain}
\begin{align}
\label{eq:Ham_transformed}
  H_{\mathrm{pl}} =&\; \sum_{\sigma=x,y,z}\sum_q\bigg\{\hbar \omega_0 {b_{q}^{\sigma}}^{\dagger} b_{q}^{\sigma} \nonumber\\
&+ \frac{\hbar \Omega}{2}  \left[ f_q^{\sigma} {b_{q}^{\sigma}}^{\dagger} \left( b_q^{\sigma} + {b_{-q}^{\sigma}}^{\dagger}\right) + \text{h.c.} \right]\bigg\}, 
 \end{align}
where the structure factor 
\begin{align}
f_{q}^{\sigma}&=2\eta_\sigma\sum_{n=1}^\infty\frac{\cos{(nqd)}}{n^3}
\nonumber \\
&=  \eta_{\sigma} \left[ \text{Li}_3 \left( \mathrm{e}^{\mathrm{i} q d} \right) 
 + \text{Li}_3 \left( \mathrm{e}^{-\mathrm{i} q d} \right) \right]
\end{align}
can be expressed in terms of the polylogarithm function $\text{Li}_s (z) = \sum_{n=1}^{\infty} z^n/n^s$. 

The Hamiltonian \eqref{eq:Ham_transformed} can be readily diagonalized by a bosonic Bogoliubov transformation, yielding
\begin{equation}
\label{eq:plchain}
 H_{\mathrm{pl}} = \sum_{\sigma=x,y,z}\sum_{q} \hbar \omega_{q}^{\sigma} {B_{q}^{\sigma}}^{\dagger} B_{q}^{\sigma},
\end{equation}
where the quasistatic spectrum of the collective plasmonic modes is
\begin{equation}
\label{eq:plspectrum}
\omega_{q}^{\sigma} = \omega_0 \sqrt{1 + 2 \frac{\Omega}{\omega_0} f_q^{\sigma}}.
\end{equation}
Notice that $-3\zeta(3)/2\leqslant f_q^\sigma/\eta_\sigma\leqslant 2\zeta(3)$, where
$\zeta(3)=\sum_{n=1}^\infty n^{-3}\simeq 1.20$ denotes Ap\'ery's constant, such that $\omega_q^\sigma$ is real for all realistic values of the ratio $d/a\geqslant 2$.

The bosonic Bogoliubov operators in Eq.\ \eqref{eq:plchain} 
are defined as $B_{q}^{\sigma} = u_{q}^{\sigma}  b_{q}^{\sigma} + v_{q}^{\sigma} {b_{-q}^{\sigma}}^{\dagger}$,
with the coefficients
$u_{q}^{\sigma} = (\omega_{q}^{\sigma} + \omega_0)/2 (\omega_0\omega_{q}^{\sigma})^{1/2}$ and $v_{q}^{\sigma} = (\omega_{q}^{\sigma} - \omega_0)/2 (\omega_0\omega_{q}^{\sigma})^{1/2}$.
The inverse transformation is $b_{q}^{\sigma} = u_{q}^{\sigma} B_{q}^{\sigma} - v_{q}^{\sigma}{B_{-q}^{\sigma}}^{\dagger}$. The operator $B_{q}^{\sigma}$ (${B_{q}^{\sigma}}^\dagger$) acts on an eigenstate $|n_q^\sigma\rangle$ of the Hamiltonian \eqref{eq:plchain} representing $n_q^\sigma$ quanta occupying the collective plasmon mode with polarization $\sigma$, wavevector $q$, and eigenenergy $\hbar\omega_q^\sigma$ with the following algebra: 
$B_{q}^{\sigma}|n_q^\sigma\rangle = \sqrt{n_q^\sigma}|n_q^\sigma-1\rangle$ 
(${B_{q}^{\sigma}}^\dagger|n_q^\sigma\rangle = \sqrt{n_q^\sigma+1}|n_q^\sigma+1\rangle$).

\begin{figure}[tb]
 \includegraphics[width=1.0\linewidth]{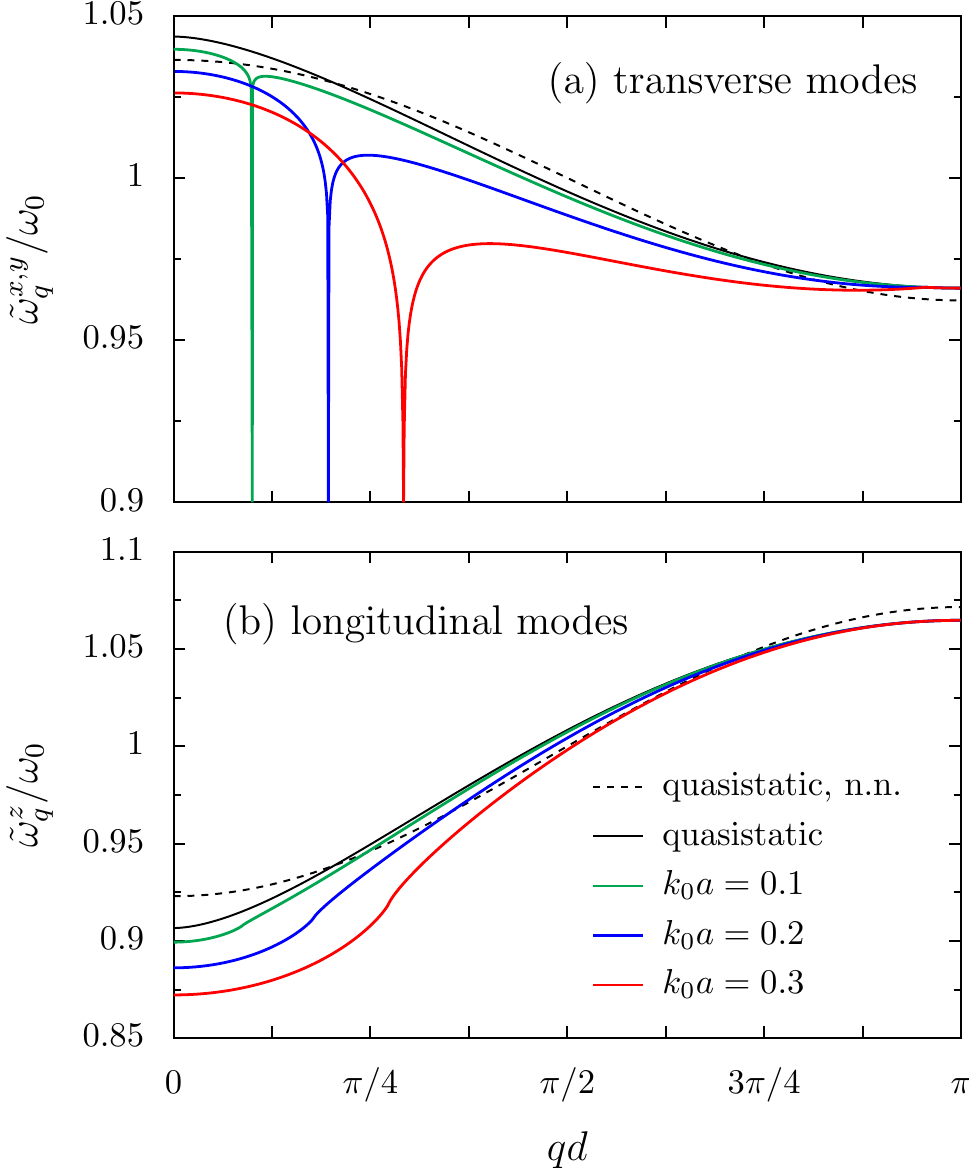}
 \caption{Collective plasmonic bandstructure (in units of the bare LSP resonance frequency $\omega_0$) as a function of the (reduced) plasmonic wavevector $q d$, in half of the first Brillouin zone. Both the (a) transverse and (b) longitudinal polarizations are shown. Solid black lines: quasistatic dispersion relation 
 without coupling to the photonic environment, see Eq.~\eqref{eq:plspectrum}. 
 Dashed lines: quasistatic dispersion relation considering only nearest-neighbor couplings from Eq.\ \eqref{eq:omega_nn}. 
 Colored lines: plasmonic dispersion relation including coupling to the photonic environment obtained from Eq.\ \eqref{renorm} with Eq.~\eqref{eq:shift_chain} for $k_0a=0.1$ (green lines), $k_0a=0.2$ (blue lines), and $k_0a=0.3$ (red lines). In the figure, the interparticle separation $d= 3 a$ (corresponding to $\Omega/\omega_0=1/54$) and the ultraviolet cutoff frequency $\omega_\mathrm{c} = c/a$.}
 \label{fig:omega}
\end{figure}

Equation \eqref{eq:plspectrum} describes the quasistatic plasmonic bandstructure of the system unperturbed by the photonic environment, which recovers the classically-calculated result from the literature \cite{Brongersma2000, Park2004, Weber2004, Citrin2004}. This is plotted in Fig.~\ref{fig:omega} (see solid black lines) for both the transverse [panel~(a)] and longitudinal polarizations [panel (b)]. 
Their and in what follows, we only display results as a function of positive plasmon wavenumber, due to the even parity of the quantities under consideration.
In the figure, we also 
plot the plasmonic dispersion considering only the dipolar interaction between nearest neighbors in the chain (see dashed lines), 
\begin{equation}
\label{eq:omega_nn}
\omega_{\textrm{n.n.},q}^\sigma=\omega_0\sqrt{1+4\eta_\sigma\frac{\Omega}{\omega_0}\cos{(qd)}}.
\end{equation}
 As can be seen from the figure, the long-ranged nature of the quasistatic part of the dipolar interaction has a rather weak effect on the plasmonic bandstructure. In particular,  
the bandwidth $\Delta\omega^\sigma=|\omega_{q=\pi/d}^\sigma-\omega_{q=0}^\sigma|$ is larger when interactions with all pairs of nanoparticles are taken into account.
In the weak-coupling limit $\Omega\ll\omega_0$, we find $\Delta\omega^\sigma=7|\eta_\sigma|\zeta(3)\Omega/2\simeq 4.21 |\eta_\sigma|\Omega$,  
while the nearest-neighbor bandwidth $\Delta\omega^\sigma_{\mathrm{n.n.}}=4|\eta_\sigma|\Omega$.

\section{Radiative shifts of the collective plasmonic bandstructure}
\label{Sec:shifts}

In order to obtain the radiative frequency shifts induced by the photonic environment
and resulting from the retardation in the dipole-dipole interaction, we now treat the coupling Hamiltonian \eqref{eq:HamCoupling} up to second order in perturbation theory. For a given mode, the plasmonic energy levels then become
$E_{n_q^\sigma} = E_{n_q^\sigma}^{(0)} + E_{n_q^\sigma}^{(1)} + E_{n_q^\sigma}^{(2)}$, where the unperturbed contribution  is $E_{n_q^\sigma}^{(0)} = n_q^{\sigma} \hbar \omega_{q}^{\sigma}$, with $\omega_{q}^{\sigma}$ as defined in Eq.~\eqref{eq:plspectrum}.

The first-order contribution 
\begin{equation}
\label{eq:E1}
E_{n_q^\sigma}^{(1)} = 2\pi \mathcal{N} \hbar\omega_0^2\frac{a^3}{\mathcal{V}}\sum_\mathbf{k}\frac{1}{\omega_\mathbf{k}}
\end{equation}
arises from the second term on the right-hand side of Eq.~\eqref{eq:HamCoupling}, that 
does not involve plasmonic degrees of freedom.  As such, this correction 
corresponds to a global energy shift which does not depend on the quantum number $n_q^\sigma$. Therefore, it does not lead to a renormalization of the collective mode resonance frequency, since only interlevel energy differences are observable.

The second-order contribution $E_{n_q^\sigma}^{(2)}$ arises from the first term on the right-hand side of the coupling Hamiltonian~\eqref{eq:HamCoupling}. 
It corresponds to the emission and subsequent reabsorption of virtual photons by the plasmonic state $|n_q^\sigma\rangle$.
Explicitly, one finds
\begin{align}
\label{eq:secondorder}
 E_{n_q^\sigma}^{(2)} =&\; \pi \hbar \omega_0^2 \omega_{q}^{\sigma} \frac{a^3}{\mathcal{V}} \sum_{\mathbf{k}, \hat{\lambda}_{\mathbf{k}}} \frac{|\hat{\sigma} \cdot \hat{\lambda}_{\mathbf{k}} |^2}{\omega_{\mathbf{k}}} \nonumber\\
 &\times
 \left(
 \frac{n_q^\sigma|F_{\mathbf{k}, q}^{-}|^2}{\omega_{q}^{\sigma} - \omega_\mathbf{k}}
 - \frac{(n_q^\sigma+1)|F_{\mathbf{k}, q}^{+}|^2}{\omega_{q}^{\sigma} + \omega_\mathbf{k}}
 \right),
\end{align}
where the summation over $\mathbf{k}$ excludes the singular term for which $\omega_{\mathbf{k}} = \omega_{q}^{\sigma}$. 
In the expression above, the array factor reads as 
\begin{equation}
\label{eq:array}
 F_{\mathbf{k}, q}^{\pm} =\frac{ \mathrm{e}^{\mp\mathrm{i} k_z  d }}{\sqrt{ \mathcal{N}}}
 \sum_{n=1}^\mathcal{N} \mathrm{e}^{ \mathrm{i} n (q \pm k_z ) d }, 
\end{equation}
with $k_z=\mathbf{k}\cdot\hat z$.
In the continuum limit, where $\sum_{\mathbf{k}} \to {\mathcal{V}} \mathcal{P}\int \mathrm{d}^3 \mathbf{k}/{(2 \pi)^3}$ (here, $\mathcal{P}$ denotes the Cauchy principal value), the second-order correction \eqref{eq:secondorder} appears to be divergent. Such a divergence can be regularized by introducing an ultraviolet cutoff $k_\mathrm{c}$, which must be of the order of $1/a$, the wavelength below which the dipolar approximation used in Eq.\ \eqref{eq:HamCoupling} breaks down \cite{footnote:renormalization}.

To second order in perturbation theory, the renormalized frequency difference between successive plasmonic energy levels $\tilde{\omega}_{q}^{\sigma} = (E_{n_q^{\sigma}+1} - E_{n_q^{\sigma}})/\hbar$ is then independent of the quantum number $n_q^{\sigma}$ and reads
\begin{equation}
\label{renorm}
\tilde{\omega}_{q}^{\sigma} = \omega_{q}^{\sigma} + \delta_{q}^{\sigma},
\end{equation}
where the radiative frequency shift is given by
\begin{equation}
\label{eqS:lambsingle}
 \delta_{q}^{\sigma} = \pi \omega_0^2 \omega_{q}^{\sigma} \frac{a^3}{\mathcal{V}} \sum_{\mathbf{k}, \hat{\lambda}_{\mathbf{k}}} 
 \frac{|\hat{\sigma} \cdot \hat{\lambda}_{\mathbf{k}} |^2}{\omega_{\mathbf{k}}}
 \left(
 \frac{|F_{\mathbf{k}, q}^{-}|^2}{\omega_{q}^{\sigma} - \omega_\mathbf{k}}
 - \frac{|F_{\mathbf{k}, q}^{+}|^2}{\omega_{q}^{\sigma} + \omega_\mathbf{k}}
 \right).
\end{equation}

Carrying out the summation over photon polarization in Eq.\ \eqref{eqS:lambsingle} via the relation 
$\sum_{\hat{\lambda}_{\mathbf{k}}}|\hat{\sigma} \cdot \hat{\lambda}_{\mathbf{k}} |^2 = 1 - ( \hat{\sigma}\cdot {\hat{k}} )^2$
and transforming the wavevector summation into a principal-value integral in spherical coordinates $(k, \theta, \varphi)$ yields 
\begin{align}
\label{eq:delta_shift}
\delta_{q}^{\sigma} =&\; \frac{1}{8 \pi^2c} \omega_0^2 \omega_{q}^{\sigma} a^3~\mathcal{P}\int_0^{k_\mathrm{c}} \mathrm{d}k\, k \nonumber\\
& \times \int_0^\pi\mathrm{d}\theta\sin{\theta}   \left(
 \frac{|F_{\mathbf{k}, q}^{-}|^2}{\omega_{q}^{\sigma} - ck}
 - \frac{|F_{\mathbf{k}, q}^{+}|^2}{\omega_{q}^{\sigma} + ck}
 \right)\nonumber\\
& \times \int_0^{2\pi}\mathrm{d}\varphi[1-(\hat k\cdot\hat\sigma)^2],
\end{align}
where $k_\mathrm{c} > \omega_{q}^{\sigma} /c$. The integral over the azimutal angle $\varphi$ is easily evaluated with the identity
\begin{equation}
\label{eqS:int_phi}
\int_0^{2\pi}\mathrm{d}\varphi[1-(\hat k\cdot\hat\sigma)^2]=
\pi|\eta_\sigma|\left(1+\mathrm{sgn}\{\eta_\sigma\}\cos^2{\theta}\right).
\end{equation}
In the long-chain limit ($\mathcal{N}\gg1$), the subsequent integral over the polar angle $\theta$ 
is readily obtained using that 
$|F_{\mathbf{k}, q}^{\pm}|^2 \simeq  2 \pi \delta \left( [q \pm k\cos\theta] d]\right)$,
where $\delta (z)$ is the Dirac delta function. 
Carrying out the remaining integral over $k$ then gives the final result
\begin{align}
\label{eq:shift_chain}
 \delta_{q}^{\sigma} =&\; \frac{\eta_\sigma}{2} \frac{\omega_0^2}{\omega_{q}^{\sigma}} \frac{q^2 a^3}{d} \Theta \left( \omega_\mathrm{c} -  c|q|  \right) \left\{ \ln \left( \frac{\omega_\mathrm{c}}{c|q|} \right) \right.\nonumber\\
 & + \left. \frac{1}{2} \left[ 1 + \mathrm{sgn} \{ \eta_{\sigma} \} \left( \frac{\omega_{q}^{\sigma}}{cq} \right)^2 \right] \ln \left( \frac{ |(cq)^2 - {\omega_{q}^{\sigma}}^2 | }{\omega_\mathrm{c}^2 - {\omega_{q}^{\sigma}}^2} \right) \right\},
\end{align}
with $\omega_\mathrm{c} = c k_\mathrm{c}$ and where $\Theta (z)$ is the Heaviside step function. 

The frequency shift \eqref{eq:shift_chain} is only logarithmically-divergent with the cutoff $\omega_\mathrm{c}$, in analogy with the expression for the Lamb shift in atomic physics~\cite{Bethe1947, Milonni1994}. As is the case for the associated radiative damping decay rate of the system (cf.\ Eq.\ \eqref{eq:gamma} in the appendix and Ref.~\cite{Brandstetter2016}), 
the magnitude of the frequency shift \eqref{eq:shift_chain} is directly proportional to the volume of the nanoparticles in the chain ($\delta_{q}^{\sigma} \propto a^3$).
When compared to the single-nanoparticle LSP radiative shift $\delta_0\simeq 2\omega_0(k_0a)^4/3\pi$ \cite{Downing2017}, the shifts \eqref{eq:shift_chain} can be at least an order of magnitude larger than $\delta_0$, since $|\delta_q^\sigma|/\delta_0\sim(k_0a)^{-1}(k_0d)^{-1}$, where $k_0a\ll1$ within the dipolar approximation \eqref{eq:HamCoupling}. Such a superradiant behavior arises due to the constructive interferences between the electromagnetic fields generated by each LSP in the chain, hence enhancing the effective coupling to the photonic environment, analogously to the cooperative Lamb shift in atomic physics \cite{Friedberg1973, Scully2009, Garrett1990, Rohlsberger2010, Keaveney2012, Meir2014}. Superradiance is also observed for the radiative decay rates of the collective plasmons \cite{Brandstetter2016} (see Fig.\ \ref{fig:gamma}), since both energy-level renormalizations and finite lifetimes are intimately related through the fluctuation-dissipation theorem \cite{Kubo1966}.

In Fig.~\ref{fig:omega}, we plot the renormalized plasmonic bandstructure \eqref{renorm} for both the transverse [panel (a)] and longitudinal polarizations [panel (b)] as solid colored lines, for the nanoparticle sizes $k_0a=0.1$ (green lines), $k_0a=0.2$ (blue lines), and $k_0a=0.3$ (red lines). 
In the figure, 
the interparticle separation is $d= 3 a$ and the ultraviolet cutoff frequency is chosen as $\omega_\mathrm{c} = c/a$. 

\begin{figure*}[tb]
 \includegraphics[width=.98\linewidth]{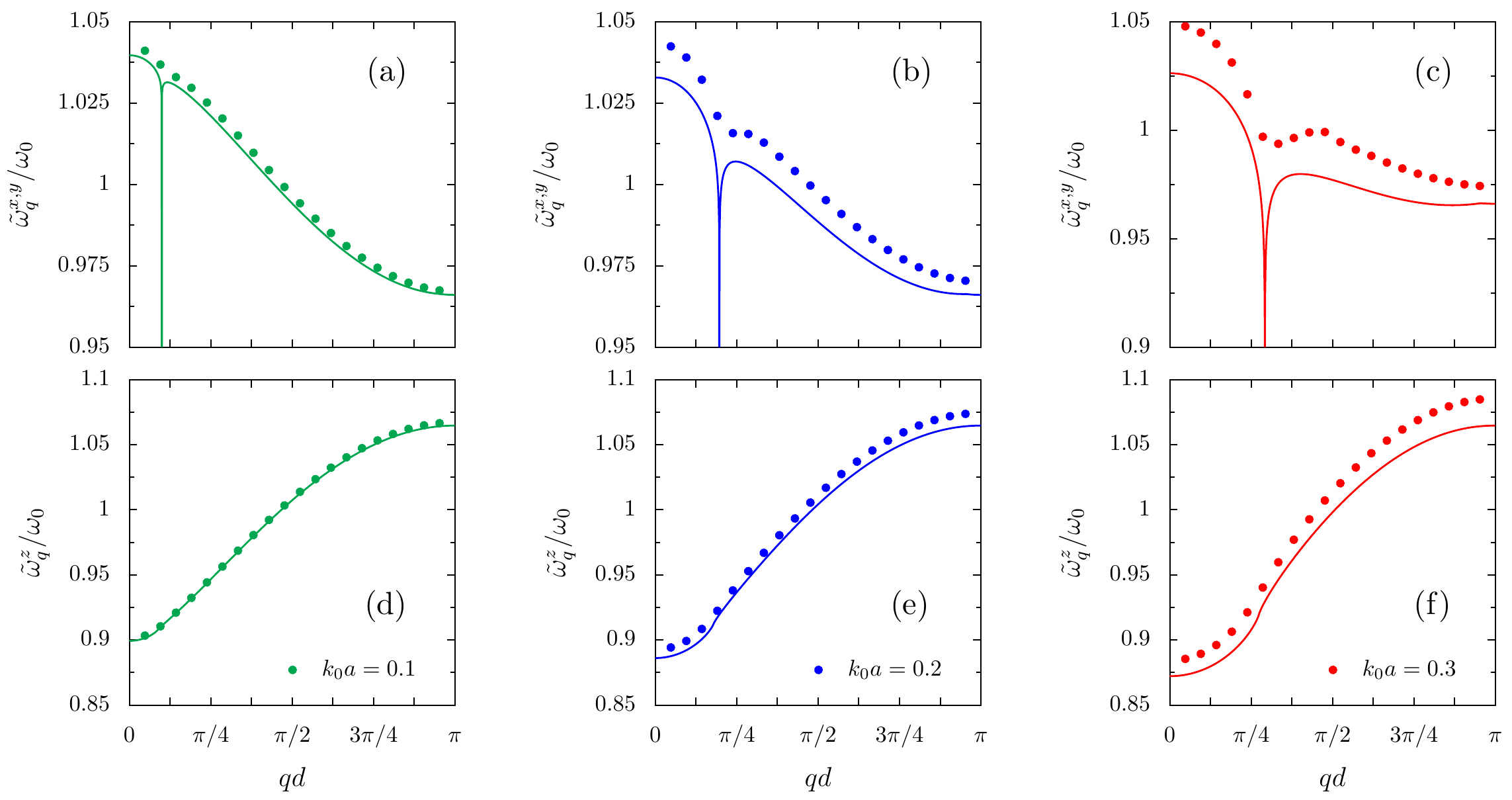}
 \caption{Plasmonic dispersion relation as a function of the wavevector for (a)-(c) the transverse and (d)-(f) 
 longitudinal polarizations. 
 Dots: fully-retarded numerical solution to Maxwell's equations for particle sizes (a),(d) $k_0a=0.1$ (green dots), (b),(e) $k_0a=0.2$ (blue dots), and (c),(f) $k_0a=0.3$ (red dots) for a chain of $\mathcal{N}=20$ nanoparticles.
Solid lines: analytical dispersion relation \eqref{renorm} for (a),(d) $k_0a=0.1$ (green lines), (b),(e) $k_0a=0.2$ (blue lines), and (c),(f) $k_0a=0.3$ (red lines). Same parameters as in Fig.~\ref{fig:omega}.}
 \label{fig:numerics}
\end{figure*}

Immediately apparent from Fig.\ \ref{fig:omega} is the presence of a cusp in the dispersion relation of the transverse modes [panel (a)]. In the large-chain limit ($\mathcal{N}\gg1$), such a cusp corresponds to a logarithmic singularity, see the last term in Eq.\ \eqref{eq:shift_chain}. This singularity occurs at the intersection between the quasistatic plasmonic dispersion and the light cone ($\omega_{q}^{x, y} = c |q|$), i.e., at $|q| \simeq  k_0$ to zeroth order in $\Omega/\omega_0\ll1$. The presence of a polarization-dependent singularity may be expected from the knowledge of the radiative damping decay rate of the system [cf.\ Eq.\ \eqref{eq:gamma} and Fig.\ \ref{fig:gamma}(a) in the appendix] which presents a step discontinuity for the transverse polarization at $|q| \simeq k_0$. In contrast, there is no such discontinuity of the radiative damping rate for the longitudinal polarization, and hence no cusp appears in the associated dispersion [see Figs.~\ref{fig:omega}(b) and \ref{fig:gamma}(b)]. 

The behavior observed in Fig.\ \ref{fig:omega} and encapsulated in Eq.\ \eqref{eq:shift_chain} has been reported previously by means of laborious numerical calculations based on the fully-retarded solutions to Maxwell's equations \cite{Weber2004, Citrin2004, Simovski2005, Citrin2006, Koenderink2006, Markel2007, Fung2007, Petrov2015}. Our simple open quantum system approach provides a transparent analytical expression which describes all of the key phenomena observed in the collective plasmon dispersion relation, including the effects of retardation. 

As a caveat, we have to point out that the singularity in the transverse collective plasmon dispersion stems from a perturbative calculation up to second 
order in the light-matter interaction. As such, large deviations from the natural LSP frequency $\omega_0$ should be treated with caution. A
thorough analysis involving the simultaneous diagonalization of the plasmonic and photonic systems in the strong coupling regime goes however beyond the scope of the present manuscript.

\section{Comparison to classical electrodynamic calculations}
\label{sec:numerics}

In this section we compare the plasmonic dispersion relation as derived from our open quantum system approach [cf.\ Eqs.\ \eqref{renorm} and \eqref{eq:shift_chain}] to the results obtained by solving Maxwell's equations including retardation. We employ the classical theory of Weber and Ford \cite{Weber2004} for a finite chain of $\mathcal{N}$ nanoparticles within the point-dipole approximation. 
Following the methodology presented in Sec.\ III of Ref.\ \cite{Weber2004}, we numerically calculate the plasmonic dispersion using the classical expression of the induced dipole moment on each nanoparticle in the chain for lossless metals. These coupled dipole moments give rise to an $\mathcal{N}\times\mathcal{N}$ non-Hermitian eigenvalue problem. The plasmonic eigenfrequencies are extracted from the real part of the complex roots of the formed determinant, while the imaginary part gives access to the radiative decay rate of the collective plasmons (see the appendix)~\cite{footnote:BC}. 

In our numerical calculations, we account for the effect of radiation damping in the polarizability using the prescription used, e.g., in Ref.\ \cite{Wokaun1982} (see also Eq.\ (6) in Ref.~\cite{Weber2004}). Notably, this ad hoc correction to the polarizability is not universally accepted in the literature and can lead to acausal behavior of the system \cite{Jackson}. Both of these problems are increasingly pronounced for greater particle sizes as this corrective term is no longer perturbative.  

In Fig.\ \ref{fig:numerics} we plot the results obtained from the numerical procedure described above for a chain of $\mathcal{N}=20$ nanoparticles, for an interparticle separation $d=3a$ (colored dots) for both polarizations and for increasing nanoparticle sizes \cite{footnote:finite-size}. We further plot the analytical expression \eqref{renorm} for comparison (see solid lines). 
As can be seen from the figure, for the smaller nanoparticle sizes [$k_0a=0.1$, green dots and lines in panels (a),(d)], the quantitative agreement is excellent for both plasmon polarizations. With increasing particle sizes [$k_0a=0.2$ and $k_0a=0.3$, blue and red dots and lines, panels (b),(e) and (c),(f), respectively], the quantitative agreement is reduced, but still very good, with a relative difference of about $\unit[1]{\%}$ ($\unit[2]{\%}$) for $k_0a=0.2$ ($k_0a=0.3$). In the transverse polarization, the frequency softening induced by the light-matter coupling (encoded in the singular frequency shift) is qualitatively reproduced by the numerical analysis.

\section{Retardation effects on the propagation of plasmonic excitations along the chain}
\label{sec:propagation}

Having discussed the influence of retardation on the plasmonic bandstructure in Sec.\ \ref{Sec:shifts}, 
here we study its impact on the propagation of plasmonic 
excitations along the chain. 
Assuming that the first nanoparticle within the chain is driven by a monochromatic, sinusoidal electric field with driving frequency 
$\omega_\mathrm{d}$, amplitude $E_0$, and polarization $\hat\epsilon$, the time-averaged, root-mean-square (dimensionless) dipole moment on nanoparticle $n$ is given by
(see Sec.\ IV in Ref.\ \cite{Brandstetter2016} for details)
\begin{equation}
\label{eq:sigma_n}
\sqrt{\Delta\sigma_n^2}=\frac{1}{\sqrt{\mathcal{N}+1}}
\sqrt{{\left(\mathcal{\tilde{S}}_n^\sigma\right)^2
+\left(\mathcal{\tilde{C}}_n^\sigma\right)^2}
}.
\end{equation}
Here, 
\begin{equation}
\mathcal{\tilde{S}}_n^\sigma=\sum_q
\mathcal{A}_q^\sigma\;
\frac{\sin{(nqd)}}
{\sqrt{\omega_q^\sigma/\omega_0}} \;
\frac{{\Omega_q^\sigma}^2-\omega_\mathrm{d}^2}
{\left(\omega_\mathrm{d}^2-{\Omega_q^\sigma}^2\right)^2+\left(\gamma_q^\sigma\omega_\mathrm{d}\right)^2}
\end{equation}
and
\begin{equation}
\mathcal{\tilde{C}}_n^\sigma=-\sum_q
\mathcal{A}_q^\sigma\;
\frac{\sin{(nqd)}}
{\sqrt{\omega_q^\sigma/\omega_0}}\;
\frac{\gamma_q^\sigma\omega_\mathrm{d}}
{\left(\omega_\mathrm{d}^2-{\Omega_q^\sigma}^2\right)^2+\left(\gamma_q^\sigma\omega_\mathrm{d}\right)^2},
\end{equation}
where
\begin{equation}
\label{eq:A}
\mathcal{A}_q^\sigma=-2\sqrt{\frac{2}
{\mathcal{N}+1}}(\hat{\sigma}\cdot\hat{\epsilon})\sin{(qd)}\, \Omega_\mathrm{R}
\tilde\omega_q^\sigma\sqrt{\frac{\omega_0}{\omega_q^\sigma}}, 
\end{equation}
with the Rabi frequency 
$\Omega_\mathrm{R}=eE_0\sqrt{N_\mathrm{e}/2m_\mathrm{e}\hbar\omega_0}$, and 
$(\Omega_q^\sigma)^2=(\tilde\omega_q^\sigma)^2+(\gamma_q^\sigma/2)^2$. 
Here, $\gamma_q^\sigma=\gamma_q^{\sigma, \mathrm{r}}+\gamma^\mathrm{O}$ corresponds to the total decay rate of the plasmonic mode with wavevector $q$ and polarization $\sigma$. 
This quantity is composed of the radiative decay rate $\gamma_q^{\sigma, \mathrm{r}}$, whose expression is given by Eq.~\eqref{eq:gamma} in the appendix, and of the nonradiative (mode- and polarization-independent) Ohmic losses $\gamma^\mathrm{O}$. 
Notice that for very small nanoparticles (those with radii smaller than ca.\ $\unit[5]{nm}$), which we do not consider in this work, one must add to $\gamma_q^\sigma$ the nonradiative Landau damping $\gamma_q^{\sigma, \mathrm{L}}$ elucidated in Ref.\ \cite{Brandstetter2016} in order to fully take into account all of the various decay mechanisms the plasmons are subject to.

\begin{figure*}[tb]
 \includegraphics[width=.88\linewidth]{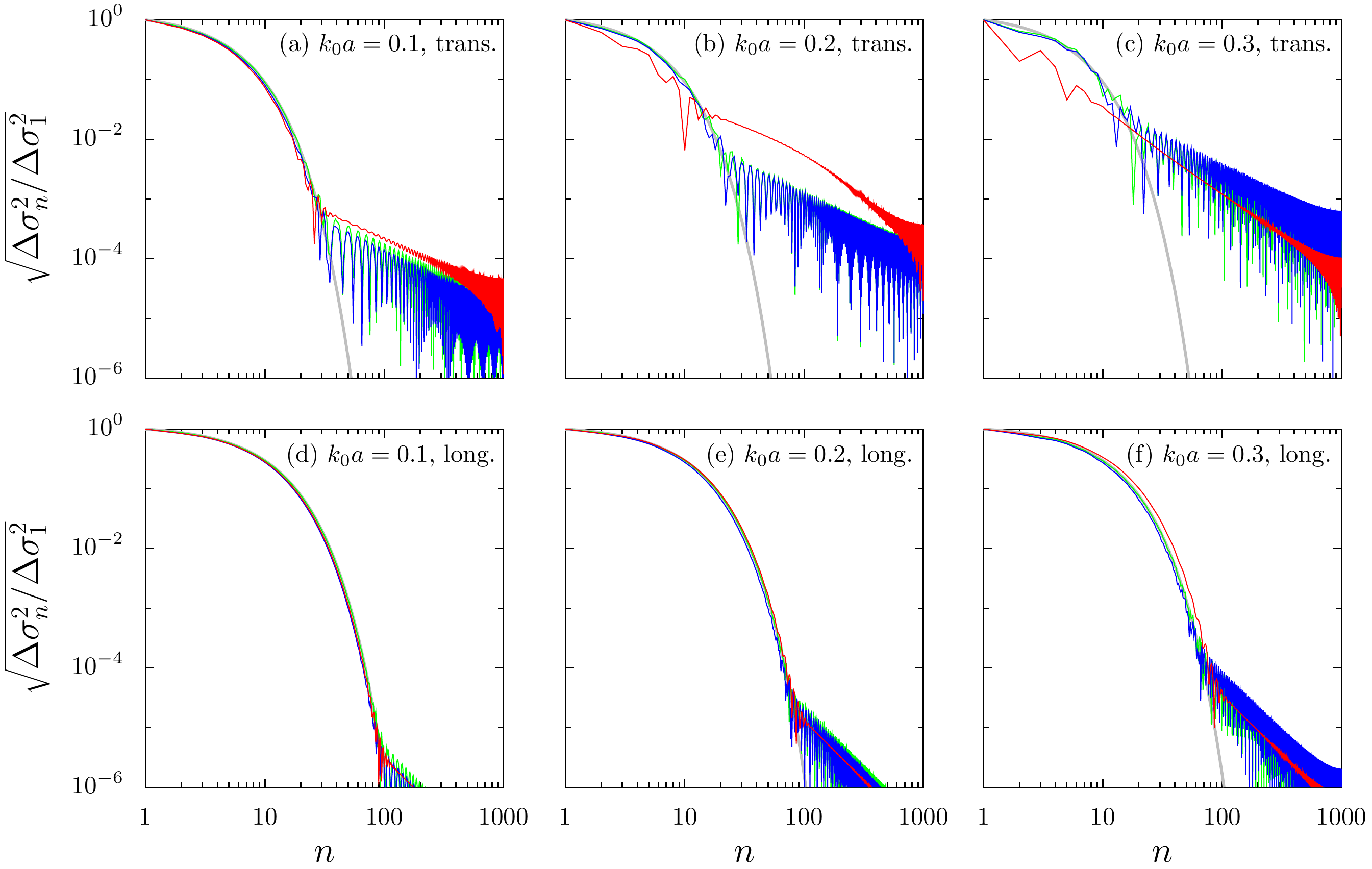}
 \caption{Normalized time-averaged root-mean-square dipole moment on nanoparticle $n$ resulting from a monochromatic excitation at frequency $\omega_\mathrm{d}=\omega_0$ of the first nanoparticle in a chain with $\mathcal{N}=1000$ and interparticle separation $d=3a$. Results are shown for 
 (a)-(c) the transverse and (d)-(f) longitudinal polarizations, and for increasing values of $k_0a$.
Red lines: full solution \eqref{eq:sigma_n}, including the radiative shifts \eqref{eq:shift_chain}.
Blue lines: solution \eqref{eq:sigma_n}, neglecting the radiative shifts \eqref{eq:shift_chain}.
Green lines: Eq.\ \eqref{eq:sigma_n}, neglecting the radiative shifts \eqref{eq:shift_chain} and with the nearest-neighbor dispersion \eqref{eq:omega_nn} instead of 
 Eq.\ \eqref{eq:plspectrum}. 
Gray lines: analytical estimate from Eq.\ \eqref{eq:sigma_n_analytical}.
 In the figure, $\gamma^\mathrm{O}=0.02\, \omega_0$.}
 \label{fig:propagation}
\end{figure*}

In Fig.\ \ref{fig:propagation}, we show the result of a numerical evaluation of Eq.\ \eqref{eq:sigma_n} (normalized by the root-mean-square dipole moment of the first nanoparticle, $\sqrt{\Delta\sigma_1^2}$) for a chain of $\mathcal{N}=1000$ nanoparticles spaced by a center-to-center interparticle distance $d=3a$, where the first particle in the chain is driven by a monochromatic field with frequency $\omega_\mathrm{d}=\omega_0$, for both transverse [panels (a)-(c)] and longitudinal [panels (d)-(f)] excitations, and for increasing values of the parameter $k_0a$. 
In the figure, the red lines correspond to the fully-retarded results, i.e., taking into account both the quasistatic dispersion relation \eqref{eq:plspectrum} including all neighbors in the chain, as well as the radiative shifts \eqref{eq:shift_chain}. The blue lines are the resulting $\sqrt{\Delta\sigma_n^2}$ 
considering the quasistatic bandstructure \eqref{eq:plspectrum} but neglecting the radiative shifts \eqref{eq:shift_chain}, while the green lines correspond to taking into account the dispersion relation with nearest-neighbor interactions only [cf.\ Eq.~\eqref{eq:omega_nn}]. The latter results correspond to the approximation used in Ref.~\cite{Brandstetter2016}. 

It is clear from Figs.\ \ref{fig:propagation}(d)-(f) that both the long-ranged nature of the quasistatic dipolar interaction and the radiative shifts have essentially no qualitative effect on the propagation of plasmons for longitudinally-polarized excitations. As was argued in Ref.\ \cite{Brandstetter2016}, there is a clear crossover between an exponentially-decaying behavior of $\sqrt{\Delta\sigma_n^2}$ for short distances along the chain ($n\lesssim80-100$ for the parameters used in the figure, depending on the nanoparticle sizes), and an algebraic one for larger distances (with $\sqrt{\Delta\sigma_n^2}\sim1/n^{\zeta^{z}}$, where $\zeta^z\simeq 2$).
The exponential decay is of purely nonradiative origin and is in excellent quantitative agreement with the estimate \cite{Brandstetter2016}
\begin{equation}
\label{eq:sigma_n_analytical} 
\sqrt{\Delta\sigma_n^2}\simeq
\frac{|\hat\sigma\cdot\hat\epsilon|}{\sqrt{2}|\eta_\sigma|}\frac{\Omega_\mathrm{R}}{\Omega}
\left[
\sqrt{1+\left(\frac{\gamma^\mathrm{O}}{4|\eta_\sigma|\Omega}\right)^2}-\frac{\gamma^\mathrm{O}}{4|\eta_\sigma|\Omega}
\right]^n, 
\end{equation}
shown by a gray line in Figs.\ \ref{fig:propagation}(d)-(f), and 
resulting in a propagation length 
$\xi^\sigma=d/\mathrm{arcsinh}(\gamma^\mathrm{O}/4|\eta_\sigma|\Omega)$.
We can therefore conclude from the above discussion that taking into account only the dipolar interaction among the nearest-neighbor nanoparticles in the chain and neglecting retardation effects [and the associated radiative frequency shifts \eqref{eq:shift_chain}] provides a very good qualitative and quantitative description of the propagation of plasmonic excitations along the chain for the longitudinal modes. 

We now focus on the propagation of plasmons for a transverse-polarized excitation [Figs.\ \ref{fig:propagation}(a)-(c)]. 
While both quasistatic results [i.e., including only nearest-neighbor interaction (green lines) and the full spectrum \eqref{eq:plspectrum} (blue lines)] are qualitatively similar, 
the fully-retarded results (red lines) show increasingly larger deviations from the quasistatically-calculated behavior 
for increasing nanoparticle sizes.
For small nanoparticle sizes [$k_0a=0.1$, panel (a)], the aforementioned crossover between exponential and algebraic decay (with a power law $\sqrt{\Delta\sigma_n^2}\sim1/n^{\zeta^{x,y}}$, with $\zeta^{x,y}\simeq1$ \cite{Brandstetter2016}) is still clear cut [with the 
exponential part of the decay well described by Eq.\ \eqref{eq:sigma_n_analytical}, see the gray lines in Figs.\ \ref{fig:propagation}(a)-(c)]. However, deviations from 
such an exponential decay become apparent for intermediate nanoparticle sizes [$k_0a=0.2$, compare blue and gray solid lines in Fig.~\ref{fig:propagation}(b)]. 
For larger sizes [$k_0a=0.3$, panel (c)], the decay of the plasmonic excitation shows a pronounced algebraic behavior (see red line), demonstrating the importance of retardation effects when describing such a decay for transverse-polarized modes.

\section{Conclusions}
\label{sec:conc}

Within an open quantum system approach, we have developed a transparent theory of collective plasmons coupled to vacuum electromagnetic modes in a chain of spherical metallic nanoparticles. Our analytical model describes how the plasmonic bandstructure of the system can be strikingly modified by retardation effects. Most noticeable is the band reconstruction for the case of the transverse plasmon polarization, which exhibits a cusp corresponding to a logarithmic singularity at the intersection of the plasmonic dispersion with the light cone. 
Our analytical results have been shown to be in excellent agreement with numerical solutions of Maxwell's equations in the chain. 

While two experiments \cite{Koendrick2007, Crozier2007} have succeeded in mapping some of the plasmonic dispersion, namely the part of the first Brillouin zone significantly 
inside the light cone, the experimental observation of the full bandstructure remains an outstanding challenge, which will most likely require the use of electron energy loss spectroscopy~\cite{Abajo2010}. 

We have further studied the influence of retardation effects on the propagation of plasmonic excitations along the chain, when the first nanoparticle is driven by a monochromatic electric field. 
While retardation effects have essentially no influence on the plasmonic propagation for longitudinally-polarized modes, the propagation of transverse-polarized modes changes from an exponential decay for short distances along the chain into a fully-algebraic one for all distances in the case of large nanoparticles.

Our expressions for the radiative frequency shifts due to the photonic environment provide a clear link to the cooperative Lamb shift phenomenon in atomic physics, and suggest our proposed nanoplasmonic system as a novel host of effects commonly thought to only belong to the realm of quantum electrodynamics.

\begin{acknowledgments}
We are grateful to Charlie-Ray Mann and Dietmar Weinmann for stimulating discussions. 
C.A.D.\ and G.W.\ acknowledge financial support from Agence Nationale de la Recherche (Project ANR-14-CE26-0005 Q-MetaMat). 
E.M.\ acknowledges financial support from the Leverhulme Trust (Research Project Grant RPG-2015-101), and the Royal Society (International Exchange Grant No.\ IE140367, Newton Mobility Grants 2016/R1 UK-Brazil, and Theo Murphy Award TM160190).
\end{acknowledgments}

\setcounter{equation}{0}
\renewcommand{\theequation}{A\arabic{equation}}
\section*{Appendix: Radiative decay rates}
\label{sec:rates}

The classical treatment based on Maxwell's equations with retardation for point dipoles proposed in Ref.~\cite{Weber2004}, which we follow to calculate plasmonic bandstructures numerically in Sec.\ \ref{sec:numerics}, further gives access to the radiative decay rate $\gamma_q^{\sigma, \mathrm{r}}$ of each collective mode in the chain of nanoparticles. 
For completeness, in this appendix we present our numerical data for $\gamma_q^{\sigma, \mathrm{r}}$, which we then compare to the results derived from the open quantum system approach of Ref.\ \cite{Brandstetter2016}. Both approaches are found to be in very good agreement. 

\begin{figure}[tb]
 \includegraphics[width=\linewidth]{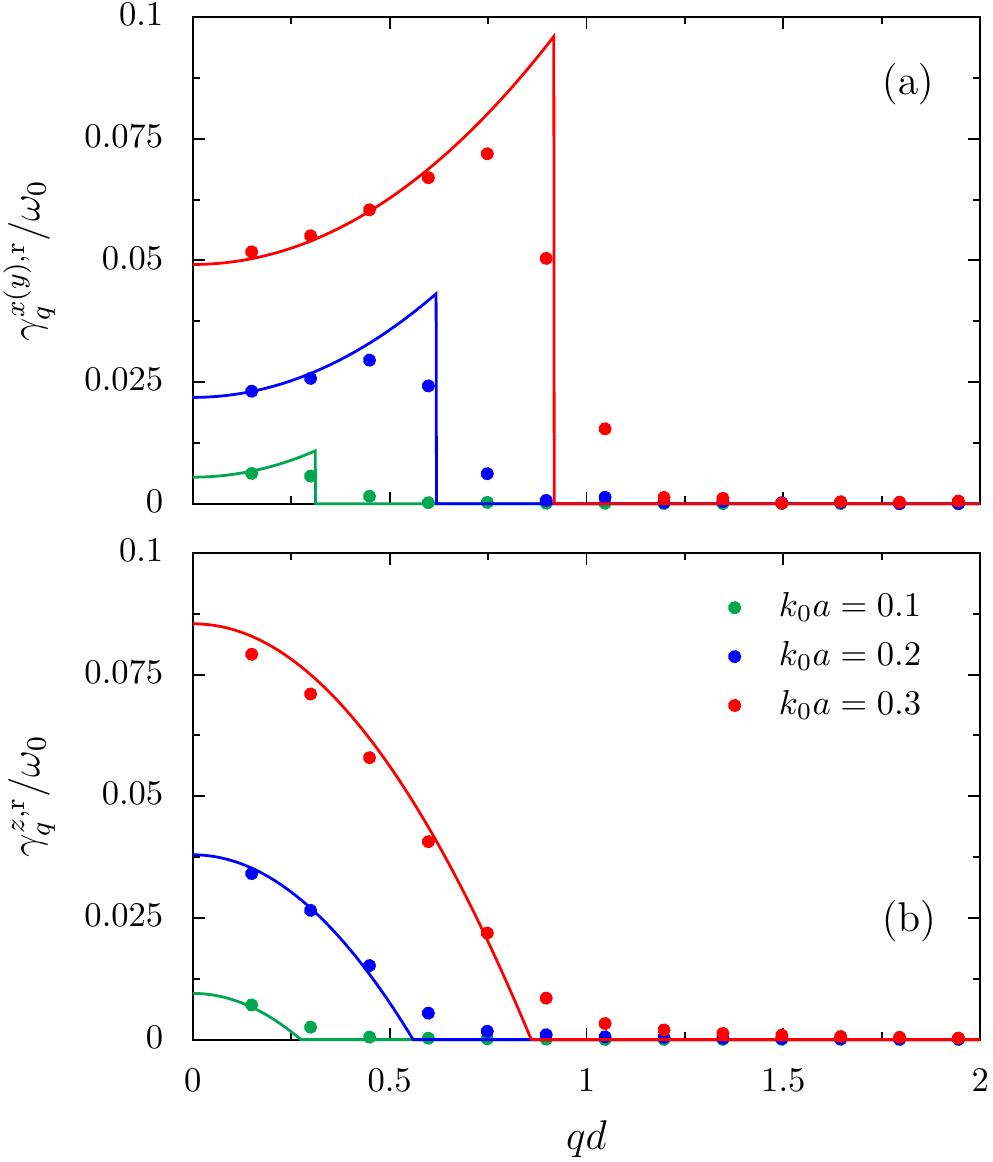}
 \caption{Radiative decay rates (in units of the resonance frequency $\omega_0$) as a function of the (scaled) wavenumber $q$ for (a) the transverse and (b) longitudinal polarizations. 
Dots: fully-retarded numerical solution to Maxwell's equations 
 for a chain of $\mathcal{N}=20$ nanoparticles separated by a distance $d=3a$
and for particle sizes $k_0a=0.1$ (green dots), $k_0a=0.2$ (blue dots), and $k_0a=0.3$ (red dots).
Solid lines: decay rate from Eq.~\eqref{eq:gamma} for $k_0a=0.1$ (green lines), $k_0a=0.2$ (blue lines), and $k_0a=0.3$ (red lines).}
 \label{fig:gamma}
\end{figure}

In Fig.\ \ref{fig:gamma}, we plot the decay rates $\gamma_{q}^{\sigma, \mathrm{r}}$ as found from our classical calculations for both collective plasmon polarizations (colored dots). As in Sec.~\ref{sec:numerics}, the presented results are for a chain of $\mathcal{N} = 20$ nanoparticles, with interparticle separation $d = 3a$ and nanoparticle sizes $k_0 a= 0.1$ (green dots), $k_0 a= 0.2$ (blue dots), and $k_0 a= 0.3$ (red dots). Also displayed by solid lines in the figure is the radiative decay rates calculated 
from Fermi's golden rule with the light-matter coupling Hamiltonian~\eqref{eq:HamCoupling}. 
Generalizing the results of Ref.\ \cite{Brandstetter2016}, which were obtained considering only nearest-neighbor interactions, to the case where the whole long-ranged quasistatic interaction is considered, yields in the long-chain limit ($\mathcal{N}\gg1$)
\begin{equation}
\label{eq:gamma}
 \gamma_{q}^{\sigma, \mathrm{r}} = \frac{\pi\eta_\sigma}{2} 
 \frac{\omega_0^2}{\omega_{q}^{\sigma}} 
 \frac{q^2 a^3}{d} \Theta \left( \omega_q^\sigma -  c|q|  \right) 
\left[ 1 + \mathrm{sgn} \{ \eta_{\sigma} \} \left( \frac{\omega_{q}^{\sigma}}{cq} \right)^2 \right].
\end{equation}
Notice that the above expression corresponds to Eq.\ (33) in Ref.\ \cite{Brandstetter2016} after the replacement of
$\omega_{\mathrm{n.n.},q}^\sigma$ [cf.\ Eq.\ \eqref{eq:omega_nn}] with $\omega_q^\sigma$ [cf.\  Eq.\ \eqref{eq:plspectrum}]. 
As can be clearly seen from Fig.~\ref{fig:gamma}, the trend encapsulated in Eq.\ \eqref{eq:gamma} is well matched by the data points. 
The slight deviations between analytical and numerical results around $q=k_0$ arise due to finite-size effects \cite{Brandstetter2016} and are most prominent for the transverse
polarization [panel (a)], as is the case for the renormalized plasmonic dispersion (see Fig.\ \ref{fig:numerics}).



\begin{thebibliography}{}
\bibitem{footnote:finite-size}
We have checked that increasing $\mathcal{N}$ does not affect the trend of the data points displayed in Fig.\ \ref{fig:numerics} so that finite-size effects are negligible, aside from the nonperturbative region for the transverse polarization where the cusp becomes increasingly noticeable for increasing chain length.
\end{thebibliography}
\end{document}